# Maskless and targeted creation of arrays of colour centres in diamond using focused ion beam technology

Margarita Lesik[1,2], Piernicola Spinicelli[1], Sébastien Pezzagna[*,1], Patrick Happel[1], Vincent Jacques[2], Olivier Salord[3], Bernard Rasser[3], Anne Delobbe[3], Pierre Sudraud[3], Alexandre Tallaire[4], Jan Meijer[5] and Jean-François Roch[2]

[1] RUBION, Ruhr-Universität Bochum, Universitätsstraße 150, 44780 Bochum, Germany
[2] Laboratoire Aimé Cotton, CNRS, Université Paris Sud and ENS Cachan, F-91405 Orsay, France
[3] Orsay Physics S. A., 95 avenue des Monts Auréliens, 13710 Fuveau, France
[4] Laboratoire des Sciences des Procédés et des Matériaux, CNRS and Université Paris 13, F-93340 Villetaneuse, France
[5] Department of Nuclear Solid State Physics, Universität Leipzig, Linnéstr. 5, 04103 Leipzig, Germany

**Keywords** Diamond, ion implantation, nitrogen-vacancy, focused ion beam

The creation of nitrogen-vacancy centres in diamond is nowadays well controlled using nitrogen implantation and annealing. Although the high-resolution placement of NV centres has been demonstrated using either collimation through pierced AFM tips or masks with apertures made by electron beam lithography, a targeted implantation into pre-defined structures in diamond may not be achieved using these techniques. We show that a beam of nitrogen ions can be focused to approximately 100 nm using focused ion beam (FIB) technology. The nitrogen ion beam is produced using an electron cyclotron resonance (ECR) plasma source. Combined with a scanning electron microscope, the nitrogen-FIB offers new possibilities for the targeted creation of single defects in diamond. This maskless technology is suitable for example for the creation of optical centres in the cavities of photonic crystals or in diamond tips for scanning magnetometry. Finally, we further show that by changing the gas source from nitrogen to xenon, standard FIB milling capabilities are also available within the same tool.

The nitrogen-vacancy (NV) defect centre in diamond is nowadays widely used for applications as magnetic sensor [1-5] and single-photon source [6-8]. It is also a promising solid-state qubit for quantum information processing [9,10], due to its unique optical and spin properties at room temperature [11,12]. The NV defect consists of a substitutional nitrogen atom and a neighbouring carbon vacancy in the diamond lattice. Generally abundant in natural diamonds [13], the NV centres can be artificially produced in synthetic material by different means. Diamonds containing substitutional nitrogen can be irradiated (with electrons, protons...) to create vacancies and then annealed, typically above 800°C, to form the nitrogen-vacancy defects [14,15]. NV centres can also be produced during the diamond growth by nitrogen doping [16]. Nevertheless, none of these two methods allows a three dimensional engineering of single NV centres. The implantation of nitrogen ions enables however such an engineering. First experiments have been realised using a nitrogen ion beam of 2 MeV energy, focused with a superconducting lens [17] yielding a resolution of approximately 300 nm on the target. Ion straggling is however an additional intrinsic limitation for the implantation resolution of high energy ions. Low energy ion implantation (typically a few keV) ensures a better ion positioning due to a limitation of the ion straggling in the range of a few nanometers, as shown by SRIM ("The Stopping Range of Ions in Matter" [18]) simulation. The high-resolution placement of single NV centres has been demonstrated with a precision better than 20 nm, using a pierced AFM tip used to collimate and address a 5 keV nitrogen beam [19]. In order to produce large-scale arrays of NV centres, a broad nitrogen implantation can be applied onto diamond which needs therefore to be masked with photoresist patterned using electron beam lithography [20,21] or covered with a mica foil containing nano- channels [22]. Although providing high spatial resolution, these techniques hardly enable the targeted creation of NV centres into some pre-defined structures on diamond.

In this article, we show that a maskless and targeted creation of arrays of NV centres is possible with focused ion beam (FIB) technology using a newly developed plasma source. Combined with a scanning electron microscope, this focused nitrogen beam offers new possibilities for nitrogen implantation and ion milling as well as easiness and rapidity. Many of the applications based on diamond and requiring single or a few NV centres to be placed within a small size region may take benefit of this new technique (figure 1).

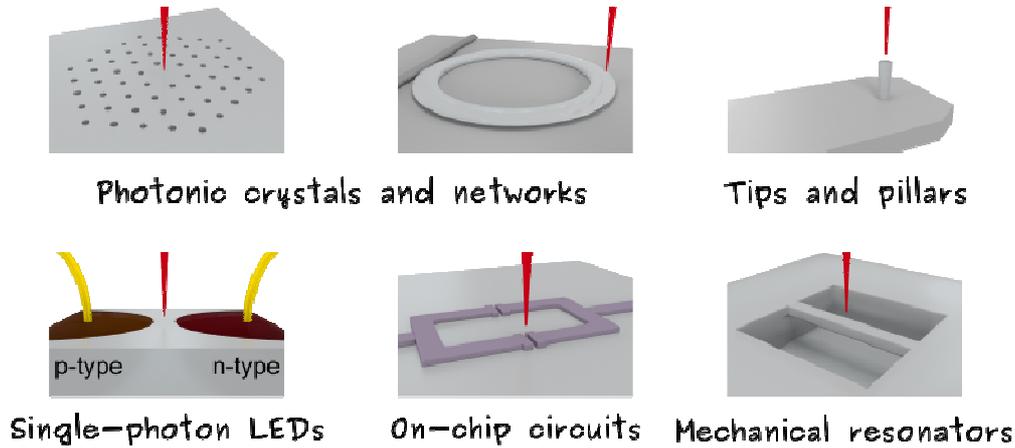

**Figure 1** Examples of applications requiring high precision placement of single NV centres, which can be reliably and reproducibly enabled using the nitrogen-FIB combined with a scanning electron microscope.

It is well suited for example to implant NV centres into diamond tips used for scanning magnetometry [5]. For photonics applications, there is also a need to place single NV centres into optical micro-cavities such as pillars [23] or photonic bandgap structures [24], as well as into waveguides [25], under a solid immersion lens [26] or at any place of a photonic network [27]. NV centres can also be easily and precisely placed in the intrinsic region of p-i-n diode structures [28,29] for single-photon emission, and also into opto-mechanical resonators [30], plasmonic structures [31], and circuits imprinted on diamond aiming at the coupling between a shallow NV centre and a superconducting qubit [32,33]. In this purpose, we produced test arrays of focused nitrogen spots to engineer NV centres and also marking crosses (10 µm long and high fluence implantation) which are visible both in electron microscopy and optical fluorescence microscopy. The implantation setup is first described. The implanted arrays of NV centres are then characterised and the resolution of the ion implantation is estimated by means of ground state depletion (GSD) microscopy [34].

**Experimental details**

The focused nitrogen beam is provided by a i-FIB column developed by Orsay Physics S.A. (figure 2a). The high-performance plasma source of ECR type (electron cyclotron resonance) [35] is mounted on a customized FIB column and used to produce the nitrogen ions (figure 2b). The FIB is installed on a scanning electron microscope chamber (FERA, Tescan) in a dual beam configuration in which the electron beam allows for fine positioning and pre-selection of structures for the focused ion beam implantation (figure 2a). Ion milling is also possible by changing the gas source from nitrogen to xenon. In these preliminary experiments, the FIB was not yet equipped with a mass filter at the moment of the implantation. The main ion species is the molecular $^{14}N_2^+$, however the beam contains a few percent of other kinds of ions, mainly $^{14}N^+$.

The diamond sample used in the following consists of an ultrapure CVD layer (200 µm thick) grown on a (001) HPHT substrate [36]. The surface of the CVD layer has been polished before ion implantation. The nitrogen concentration in the CVD layer is in the range of $2\times10^{14}$ cm$^{-3}$. For the ion implantation, the acceleration voltage has been chosen to be 30 kV.

Several arrays of focused nitrogen spots (21×21 with 2 µm separation) have been implanted (figure 2c), with each, a constant implantation time per spot (from 30 µs to 1 ms). With a constant beam current of 0.3 pA, this corresponds to a number of implanted nitrogen atoms per spot between 110 and 3700. It is expected that 1 to 5 % of the implanted nitrogen will be converted into NV centres after annealing the diamond two hours at 800°C in vacuum [37].

The depth of the N-implantation at 15 keV energy per atom can be estimated by SRIM simulation to be around 20 nm. In order to remove contamination or graphite which may have been produced at the diamond surface during annealing (and then lead to parasitic effects such as photoluminescence quenching), the sample has been cleaned in a boiling mixture of acids (perchloric, sulphuric and nitric 1:1:1) for 4 hours. This moreover stabilises the charge state of the shallow NV centres in their negatively charged state NV$^-$ [38,39]. Aside of each pattern, a marking cross implanted at high fluence has been drawn with the nitrogen beam. As seen in figures 2d and 3, such a mark presents the advantage of being visible both with a confocal microscope, due to the bright luminescence of this NV ensemble, and with a scanning electron microscope. This reinforces the flexibility of this ion implantation dual setup.

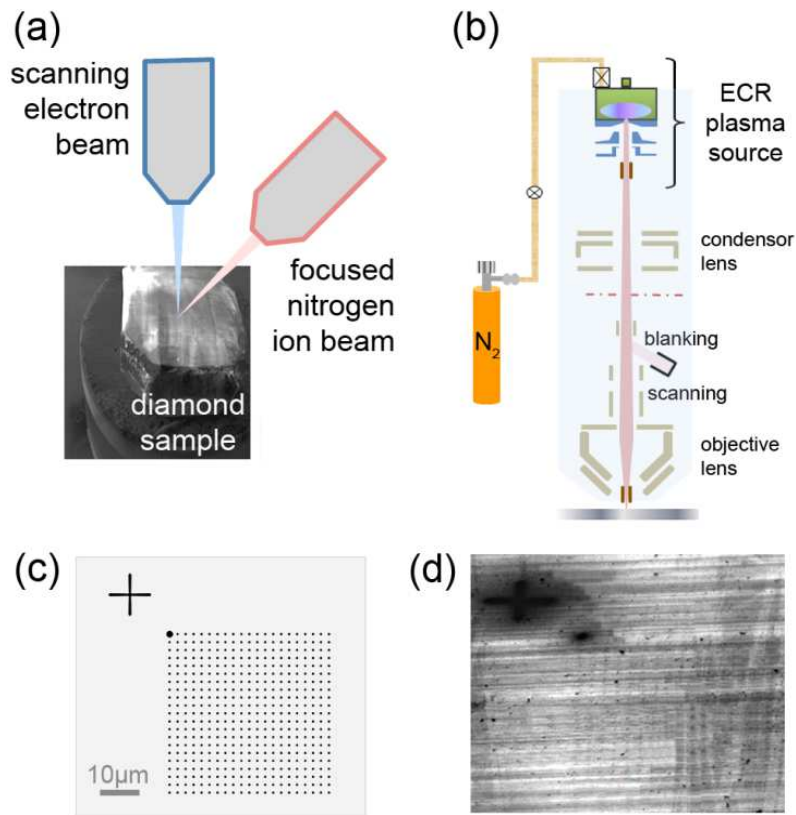

**Figure 2** (a) Dual beam used for the ion implantation experiments. (b) Scheme of the nitrogen ion column developed by Orsay Physics S.A. and based on a ECR plasma source. (c) Test pattern for the implantation with a marking cross (high fluence). (d) Scanning electron microscopy image (tilted) of the implanted diamond showing the nitrogen pattern. The high ion fluence induced amorphisation of the diamond.

The implanted structures have been characterised using a home-made scanning confocal microscope with laser excitation at 532 nm wavelength. The excitation beam is focused on the diamond sample using an oil-immersion microscope objective with a NA = 1.35 numerical aperture. The luminescence from the excited NV centres is collected with the same microscope objective and spectrally filtered with a dichroic mirror in order to remove any excitation residuals. The detection system consists of a photon counting system (two avalanche photodiodes mounted in a Hanbury Brown and Twiss configuration) and of a spectrograph based on a cooled CCD matrix. The spectral measurement is used to detect the charge state of the implanted NV centres.

## Results and discussion

As mentioned above, all the fabricated arrays were designed using the same pattern (figure 2c), which has the appearance of a grid (40 µm × 40 µm) with implanted N ions in each unit. The spots are separated by 2 µm distance from each other. The ion current was kept constant and the quantity of implanted ions was controlled by changing the dwell time. The detected photoluminescence of one of the implanted patterns (1 ms of dwell time per each spot) at an optical excitation of 150 µW is presented in figure 3. It consists of well separated bright spots each containing tens of NV centres that the confocal setup cannot optically resolve. By comparing the mean intensity recorded for a single NV centre to the average luminescence intensity emitted from these implanted spots, it is roughly found that they are made out of 30 to 40 NV centres. This corresponds to a creation yield of ~ 1%.

The measured spectra of the implanted spots in the different arrays prove that the fabricated NV centres are mostly present in their negatively charged $NV^-$ state (figure 3). It can also be seen that a relatively high and unwished density of NV centres (1 to 5 $\mu m^{-2}$) is present at the diamond surface. This is due to the fact that the implantation has been performed without any mass filter mounted on the FIB column. Although this drawback can be easily avoided using a mass separation E×B Wien filter, the effect of such a filter on the focusing properties of the FIB column will require further studies.

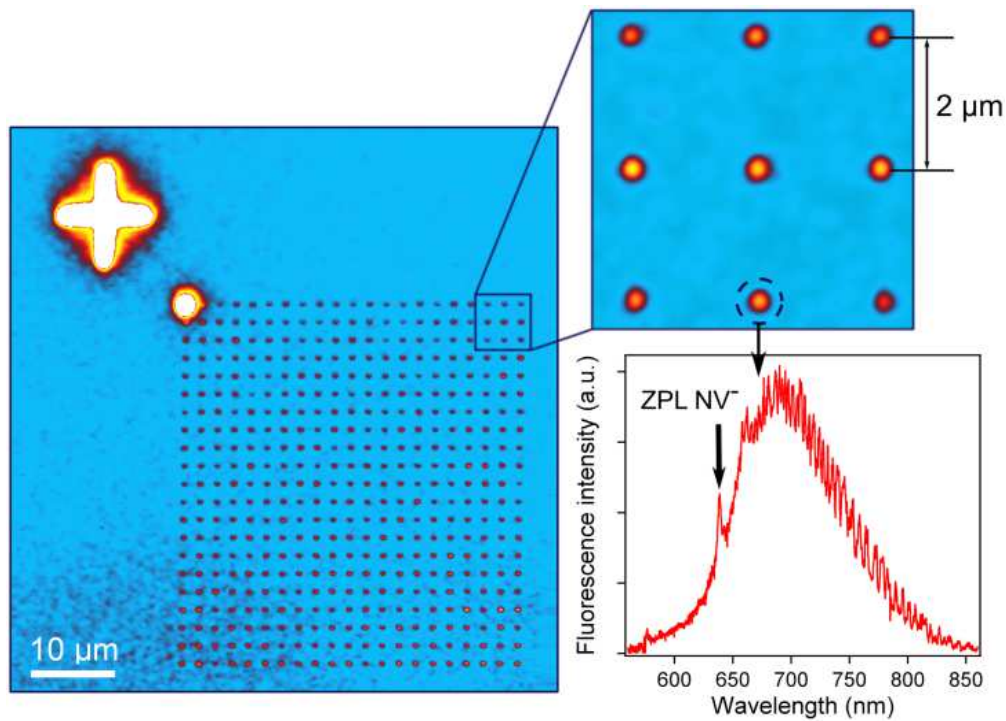

**Figure 3** Luminescence scan of a diamond surface implanted with a focused beam of $^{14}N_2^+$ ions. The inset shows a closer view of the array of NV centres and the emission spectrum of one of the implanted spots.

The resolution of the ion implantation which is determined by the focus of the ion beam, is inferred using two complementary analysis. Firstly, the widths of nine implanted spots at the higher fluence (1 ms dwell time) have been measured with confocal microscopy and compared to the width obtained for a single shallow NV centre. The single NV centre image is used to determine the point spread function (PSF) of the confocal system (figure 4a).

For a single NV, we measured a width of ~ 250 nm, in good agreement with the optical diffraction-limited resolution given by $\Delta r \approx 1.22 \times \lambda/(2 \times NA)$ (where $\lambda$ is the optical wavelength and NA the numerical aperture of the objective). By deconvolving the measured implanted spots from the PSF, it is found that the ion beam focus is on average 90 ± 20 nm (figure 4c).

The second analysis is based on high-resolution imaging of the implanted spots using Ground State Depletion (GSD) microscopy in order to determine the number of NV centres per spot and resolve their lateral distribution. In GSD, the intensity profile of the excitation beam is modulated by placing a vortex phase mask in the optical path which produces a doughnut-shaped excitation beam [34]. Increasing the excitation power leads to a reduction of the size of the dip observed in the PSF. This is shown for a single NV centre in figure 4d (excitation power of 50 mW) where an imaging resolution of 35 nm was obtained. Ideally, GSD measurements should be done on the spots implanted with the highest dwell time of 1000 µs. However, the NV density in these spots was too high to optically resolve each centre. Thus the GSD analysis has been performed on one particular spot belonging to the lower fluence array generated with a 300 µs dwell time. The confocal image of this spot is shown in figure 4b. From its fluorescence intensity and from the corresponding intensity autocorrelation function recorded with the Hanbury Brown and Twiss setup (inset in figure 4b), it can be estimated that the spot consists of $\simeq 9$ NV centres. Figure 4e shows the GSD scan of this implanted spot.

The number of NV centres and their relative positions can be estimated, as shown in figure 4f. The size of the region in which the nine NV centres are found illustrates the ion beam focus, which was here between 100 and 150 nm. This is a rough estimate due to the fact that the high surface density of NV centres surely worsens the result. The two methods used to estimate the resolution of the focused nitrogen ion beam are nevertheless in good agreement. They give evidence for a beam resolution in the order of 100 nm.

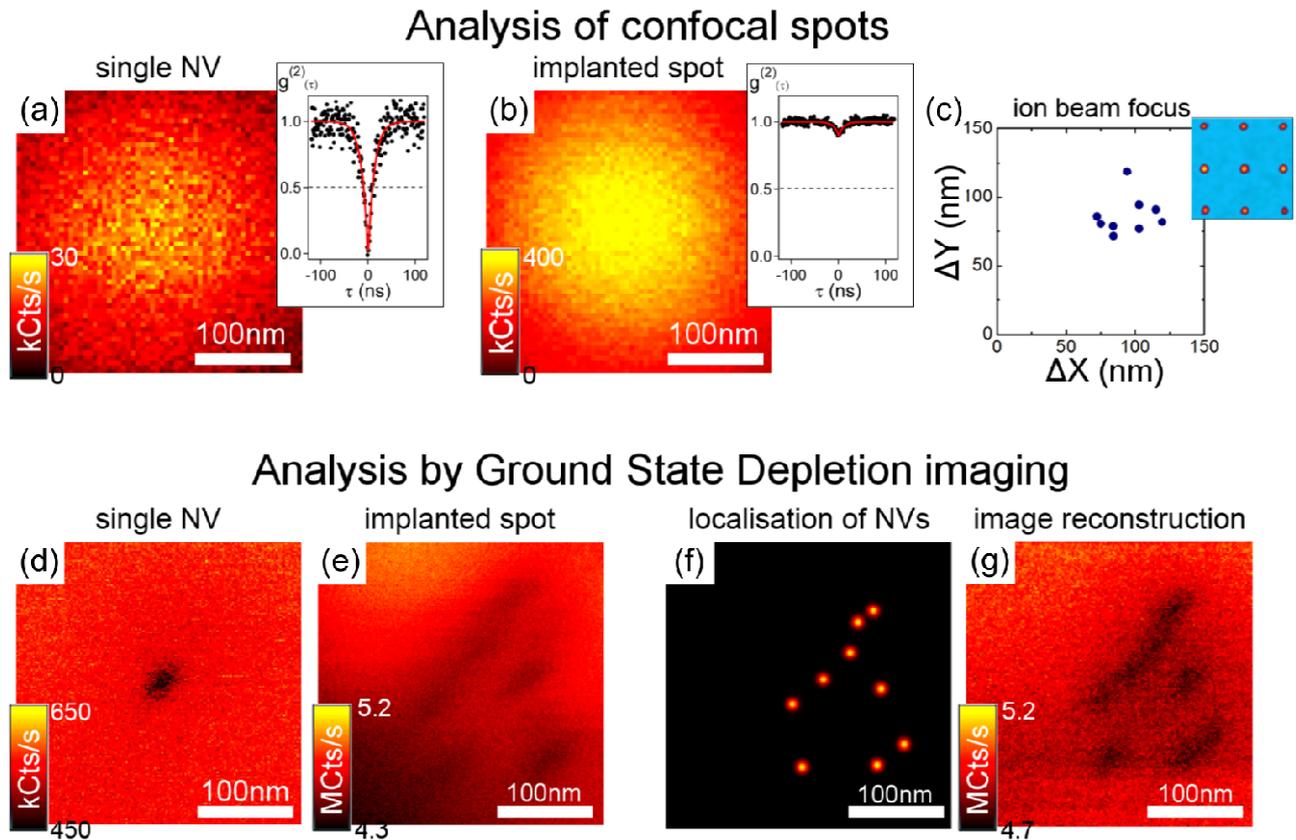

**Figure 4** Estimation of the ion beam focus. (a) and (b) Confocal scans of, respectively, a shallow implanted single NV centre and a focused implantation spot corresponding to a 300 µs dwell time, with the relative autocorrelation function (normalised raw data). The laser excitation power is 75µW in order to keep the NV centres below saturation. (c) Ion beam focus determined for the nine implanted spots shown into the inset (fluence of 1000 µs dwell time) and determined by 2D deconvolution using the confocal scan of the single NV as point spread function of the confocal setup. (d) and (e) GSD scans of, respectively, the same single NV and implantation spot as in (a) and (b). The laser excitation power is 50 mW. (f) Deduced positions (from (e)) of the implanted NVs. (g) GSD image reconstructed by convolving the matrix (f) and the experimental PSF (d).

The ECR plasma source is moreover able to generate other kinds of ions. The gas source can for example be changed from nitrogen to xenon in order to use the FIB as a milling and micro-structuration tool (and thus avoiding nitrogen contamination provided that a mass filter is used). Xe ions possess a sputter yield twice higher as gallium ions typically used for milling with FIBs using liquid metal ion sources. Micro-lenses or pillars can for example be created at the same time as the nitrogen implantation. Such a micro-pillar (10 µm diameter, 5 µm height and surrounded by a 5µm channel) has been fabricated on the same diamond sample using 30 keV $Xe^+$ ions and is shown in figure 5. Note that this milling process avoids the sample contamination with heavy ions which occurs with standard gallium FIBs.

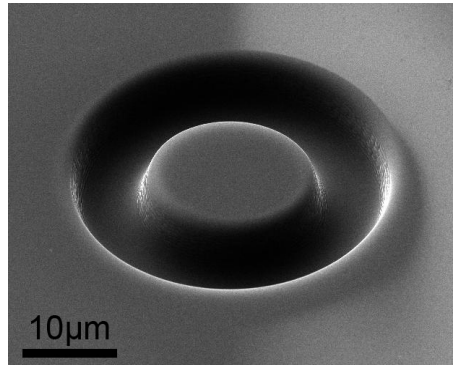

**Figure 5** Milling of a micro-structure using a focused xenon ion beam.

In conclusion, we have presented a newly developed FIB column based on an ECR plasma source which can produce a focused beam of nitrogen ions. Combined with a scanning electron microscope, the nitrogen-FIB opens new perspectives for the targeted creation of nitrogen-vacancy centres into nano- and micro-structures made in diamond for a wide range of applications. In preliminary experiments, the ion beam focus has been estimated by optical means using the luminescence emitted by the NV centres produced by the nitrogen implantation. It was found that the beam size is in the range of 100 nm. A mass filter will further be implemented on the ion beam column in order to avoid the contamination due to a rest of other ion species accelerated out of the plasma source and co-implanted with the $^{14}N_2^+$ main ion species. Finally, micro-structures can be easily milled in diamond with this FIB using xenon ions, at fast milling rate and without any contamination of the sample with heavy ions as with standard gallium FIB.

**Acknowledgements** We acknowledge fruitful discussions with Daniel Comparat who pinpointed the potential of the ECR plasma source for performing versatile ion implantation. The work is supported by the DFG (FOR1493, Diamond Materials for Quantum Application), the European union/BMBF (CHIST-ERA QINCV, Quantum Information with NV centre), the Volkswagen Stiftung and the French ANR (ADVICE project).

**References**
[1]   G. Balasubramanian *et al.*, Nature **455**, 648 (2008).
[2]   J. Maze *et al.*, Nature **455**, 644 (2008).
[3]   C. L. Degen, Appl. Phys. Lett. **92**, 243111 (2008).
[4]   L. Rondin, J.-P. Tetienne, P. Spinicelli, C. Dal Savio, K. Karrai, G. Dantelle, A. Thiaville, S. Rohart, J.-F. Roch and V. Jacques, Appl. Phys. Lett. **100**, 153118 (2012).
[5]   P. Maletinsky, S. Hong, M. S. Grinolds, B. Hausmann, M. D. Lukin, R. L. Walsworth, M. Loncar and A. Yacoby, Nature Nanotechnology **7**, 320 (2012).
[6]   C. Kurtsiefer, S. Mayer, P. Zarda and H. Weinfurter, Phys. Rev. Lett. **85**, 290 (2000).
[7]   V. Jacques, E. Wu, F. Grosshans, F. Treussart, P. Grangier, A. Aspect and J.-F. Roch, Science **315**, 966 (2007).
[8]   I. Aharonovich, S. Castelletto, D. A. Simpson, C.-H. Su, A.-D. Greentree and S. Prawer, Reports Prog. Phys. **74**, 076501 (2011).

[9] F. Dolde, I. Jakobi, B. Naydenov, N. Zhao, S. Pezzagna, C. Trautmann, J. Meijer, P. Neumann, F. Jelezko and J. Wrachtrup, Nature Physics **9**, 139 (2013).
[10] H. Bernien *et al.*, arXiv: 1212.6136 (2013).
[11] A. Gruber, A. Dräbenstedt, C. Tietz, L. Fleury, J. Wrachtrup and C. von Borczyskowski, Science **276**, 2012 (1997).
[12] F. Jelezko. T. Gaebel, I. Popa, A. Gruber and J. Wrachtrup, Phys. Rev. Lett. **92**, 076401 (2004).
[13] A. Zaitsev, *Optical properties of diamond. A data handbook*, (Springer) ISBN 3-540-66582-X (2001).
[14] Y.-R. Chang *et al.*, Nature Nanotech. **3**, 284 (2008).
[15] J. Botsoa, T. Sauvage, M.-P. Adam, P. Desgardin, E. Leoni, B. Courtois, F. Treussart and M.-F. Barthe, Phys. Rev. B **84**, 125209 (2011).
[16] K. Ohno, F. J. Heremans, L. C. Bassett, B. A. Myers, D. M. Toyli, A. C. Bleszynski Jayich, C. J. Palmstrom and D. D. Awschalom, Appl. Phys. Lett. **101**, 082413 (2012).
[17] J. Meijer, B. Burchard, M. Domhan, C. Wittmann, T. Gaebel, I. Popa, F. Jelezko and J. Wrachtrup, Appl. Phys. Lett. **87**, 261909 (2005).
[18] J. Ziegler, *The stopping range of ions in matter*, SRIM-2008, online at http://srim.org (2008).
[19] S. Pezzagna, D. Wildanger, P. Mazarov, A. D. Wieck, Y. Sarov, I. Rangelow, B. Naydenov, F. Jelezko, S. W. Hell and J. Meijer, Small **6**, 2117 (2010).
[20] D. M. Toyli, C. D. Weiss, G. D. Fuchs, T. Schenkel and D. Awschalom, Nano Lett. 10, 3168 (2010).
[21] P. Spinicelli et al., New J. Phys. 13, 025014 (2011).
[22] S. Pezzagna, D. Rogalla, H.-W. Becker, I. Jakobi, F. Dolde, B. Naydenov, J. Wrachtrup, F. Jelezko, C. Trautmann and J. Meijer, Phys. Stat. Sol. A **208**, 2017 (2011).
[23] T. M. Babinec, B. J. M. Hausmann, M. Khan, Y. Zhang, J. R. Maze, P. R. Hemmer and M. Loncar, Nature Nanotechnology **5**, 195 (2010).
[24] J. Riedrich-Möller *et al.*, Nature Nanotechnology **7**, 69 (2012).
[25] P. Olivero *et al.*, Advanced Mater. **17**, 2427 (2005).
[26] L. Marseglia *et al.*, Appl. Phys. Lett. **98**, 133107 (2011).
[27] B. J. M. Hausmann *et al.*, Nano Letters **12**, 1578 (2012).
[28] A. Lohrmann, S. Pezzagna, I. Dobrinets, P. Spinicelli, V. Jacques, J.-F. Roch, J. Meijer and A. M. Zaitsev, Appl. Phys. Lett. **99**, 251106 (2011).
[29] N. Mizuochi *et al.*, Nature Photonics **6**, 299 (2012).
[30] O. Arcizet, V. Jacques, A. Siria, P. Poncharal, P. Vincent and S. Seidelin, Nature Physics 7, 879 (2011).
[31] S. Schietinger, M. Barth, T. Aichele and O. Benson, Nano Letters **9**, 1694 (2009).
[32] Y. Kubo *et al.*, Phys. Rev. Lett. **107**, 220501 (2011).
[33] X. Zhu *et al.*, Nature **478**, 221 (2011).
[34] E. Rittweger, D. Wildanger and S. W. Hell, Europhysics Letters **86**, 14001 (2009).
[35] P. Sortais, T. Lamy, J. Medard, J. Angot, P. Sudraud, O. Salord and S. Homri, Rev. Scientific Instrum. **83**, 02B912 (2012).
[36] A. Tallaire, A. T. Collins, D. Charles, J. Achard, R. Sussmann, A. Gicquel, M. E. Newton, A. M. Edmonds and R. J. Cruddace, Diam. Relat. Mater. **15**, 1700 (2006).
[37] S. Pezzagna, B. Naydenov, F. Jelezko, J. Wrachtrup and J. Meijer, New J. Phys. **12**, 065017 (2010).
[38] B. Grotz *et al.*, Nature Communications **3**, 729 (2012).
[39] L. Rondin *et al.*, Phys. Rev. B **82**, 115449 (2010).